# Predicting the Plant Root-Associated Ecological Niche of 21 *Pseudomonas* Species Using Machine Learning and Metabolic Modeling

**Jennifer Chien (1), Peter Larsen (2)**

(1) Wellesley College, Wellesley, MA, United States of America
(2) Argonne National Laboratory, Biosciences Division, Argonne, IL, United States of America

---

## Abstract:

Plants rarely occur in isolated systems. Bacteria can inhabit either the endosphere, the region inside the plant root, or the rhizosphere, the soil region just outside the plant root. Our goal is to understand if using genomic data and media dependent metabolic model information is better for training machine learning of predicting bacterial ecological niche than media independent models or pure genome based species trees. We considered three machine learning techniques: support vector machine, non-negative matrix factorization, and artificial neural networks.

In all three machine-learning approaches, the media-based metabolic models and flux balance analyses were more effective at predicting bacterial niche than the genome or PRMT models. Support Vector Machine learning with a linear kernel trained on a minimal media base with Mannose, Proline and Valine was most predictive of all models and media types with an f-score of 0.8 for rhizosphere and 0.97 for endosphere. Thus we can conclude that media-based metabolic modeling provides a holistic view of the metabolome, allowing machine learning algorithms to highlight the differences between and categorize endosphere and rhizosphere bacteria. There was no single media type that best highlighted differences between endosphere and rhizosphere bacteria metabolism and therefore no single enzyme, reaction, or compound that defined whether a bacteria's origin was of the endosphere or rhizosphere.

There was no one media that was the most predictive across all machine-learning algorithms, indicating that there is no single function or gene that defines bacterial niche limitations. In all simulations, the media dependent metabolic models were more predictive than pure genome or media dependent data trained machine learning models. The difference between the bacterial inhabitance of either the rhizosphere or endosphere could be added functionality allowing exploitation of another niche, rather than bacteria only inhabiting one ecological niche or another.

**Keywords:**
Pseudomonas, SVM, ANN, NMF, FBA, endosphere, rhizosphere, metabolic model

---

## Background:

Plants and bacteria share a symbiotic relationship; they interact and communicate with one another. In these interactions, plants provide carbon, in the form of photosynthetically-derived sugars, to the bacterial symbiotes. The bacteria, in return, provide greater access to soil nutrients, defense against pathogens, and protection from abiotic stresses to their host plant. There are two possible ecological niches that these symbiotic bacteria might occupy: the endosphere and the rhizosphere. Bacteria associated with the endosphere inhabit plant tissue, without causing apparent harm to the host [1, 2]. Other bacteria live in the rhizosphere, where the soil is directly influenced by exudates from plant roots [2]. These ecological niches are, however, difficult to examine directly and many bacteria in either niche cannot be easily cultures in the laboratory for in depth characterization. Rather, these ecological niches can only be assayed through metagenomics techniques and direct genomic sequencing of the bacterial species found there.

Genomic information alone, however, is not always sufficient to characterize novel bacteria. In a recent publication, Support Vector Machine (SVM) learning was used to identify the most predictive molecular mechanisms indicative of ecological role in terms of biofilm, biocontrol agent, promoter of plant growth, and plant pathogen **[3]**. In this analysis, it was discovered that metabolic models constructed from genomic data were more predictive of ecological niche than was genomic information alone. While this approach was successful to predict very broad categories of ecological niches for Pseudomonads, the distinguishing features between the endosphere and the rhizosphere are expected to be much finer, requiring a more targeted metabolic modeling approach. Following on that observation, we use machine learning algorithms to identify if and what the distinguishable characteristics are between *Pseudomonas* species that inhabit the two closely related but ecologically distinct environments, endosphere and rhizosphere.

We used machine learning approaches and metabolic modeling to determine if there was any one algorithm or media that could best predict the ecological niche for *Pseudomonas* species that occupy the *Populus deltoids* microbiome. For this analysis, we considered twenty-one *Pseudomonas* species genomes from sites along the Caney Fork River in central Tennessee and at the Yadkin River in North Carolina that have recently been sequenced and specific ecological niche characterized **[4]**. These genomes were used to construct two types of metabolic models: Predicted Relative Metabolic Turnover (PRMT) and Flux Balance Analysis (FBA). PRMT, a media environment-independent modeling method effective at quantifying secondary metabolic reactions, has been used successfully in previous analysis efforts **[5]**. FBA, unlike PRMT, is a media-depended metabolic modeling method. FBA models were constructed using 14 different nutrient conditions, considering carbon sources previously identified as distinguishing endosphere and rhizosphere environments. We considered three machine learning methods: Support Vector Machine (SVM), Artificial Neural Networks (ANN), and Non-negative Matrix

Factorization (NMF). We used three sets of data to train our models: annotated genomes, PRMT metabolic models, and 15 FBA metabolic models.

The three machine learning algorithms used vary in runtime, complexity, and supervision level. Support Vector Machine Learning (SVM) is a form of supervised learning often used for data clustering and anomaly detection **[6]**. The SVM algorithm builds a model that maximizes the separation between data points in each category, with a hyperplane (default is a line, but model varies with the specified kernel). The svm() function in R package e1071 can build a model given a training data set and predict the classification of additional data points. SVM is particularly useful because it is fast and there is no danger of over-fitting to the data.

An Artificial Neural Network (ANN), modeled after the way human brains function and learn, is an algorithm that builds a model with the specified number of hidden layers with randomized weights that are adjusted, through supervised learning with training data, to optimize the outcome. The category of additional data is predicted with a numeric value between 0 and 1, indicating R and E, respectively. The neuralnet() function in the neuralnet package in R is able to account for more complex relationships in the data than SVM. However, it often takes longer to train than SVM and there is a chance of overfitting the model to the data.

Non-negative Matrix Factorization (NMF) is a form of unsupervised machine learning that groups the data into a specified number of categories. NMF is often used for discovering structure, but is also slower than SVM to train. NMF takes a matrix of non-negative values and factors it into a matrix of features and a matrix of samples. The model places all species into the number of specified groups, regardless of known ecological niche.

## Implementation:

### Selection and annotation of rhizosphere and endosphere associated *Pseudomonas*

*Pseudomonas* species occupy a diverse set of niches and roles as human pathogen, plant pathogen, and soil bacterium. There were twenty-one fully sequenced Pseudomonad strains available on NCBI (http://www.ncbi.nlm.nih.gov/genome) associated with research investigating diversity and plant-root interaction in the *Populus Deltoides* microbiome **[4]**. Pseudomonad origin was defined as a categorical variable, either endosphere or rhizosphere and assigned based on published manuscripts (**Table 1**). For genomic annotation and FBA metabolic construction, KBase (http://kbase.us) was used. The genomes were uploaded, annotated (using the "Annotate Microbial Genome" method), and renamed as Annot_*originalName* (ex. Annot_PseuGM16E) as the standard form of genome analysis. All annotated genomes were selected in the "Insert Genome Into Species Tree" method and species trees were produced for neighbor public genome count of 10, 30, 70, and 100 (ex. PseudomonasGenomeTree_#Neighbor). Each Pseudomonad genome on the tree was color coded based on its origin to determine if there was a distinguishable genomic pattern between the strains that inhabit the rhizosphere and endosphere. A pattern would indicate there is a genetic marker differentiating between in species that inhabit either niche.

| **Organism** | **Strain** | **Origin**[a] | **Accession #** |
|---|---|---|---|
| *Pseudomonas sp.* | GM16 | E | AKJV00000000 |
| *Pseudomonas sp.* | GM84 | R | AKJC00000000 |
| *Pseudomonas sp.* | GM24 | E | AKJR00000000 |
| *Pseudomonas sp.* | GM102 | E | AKJB00000000 |
| *Pseudomonas sp.* | GM33 | E | AKJO00000000 |
| *Pseudomonas sp.* | GM78 | E | AKJF00000000 |
| *Pseudomonas sp.* | GM80 | E | AKJD00000000 |
| *Pseudomonas sp.* | GM17 | E | AKJU00000000 |
| *Pseudomonas sp.* | GM79 | E | AKJE00000000 |
| *Pseudomonas sp.* | GM67 | E | AKJH00000000 |
| *Pseudomonas sp.* | GM74 | R | AKJG00000000 |
| *Pseudomonas sp.* | GM30 | E | AKJP00000000 |
| *Pseudomonas sp.* | GM41 | E | AKJN00000000 |
| *Pseudomonas sp.* | GM50 | E | AKJK00000000 |
| *Pseudomonas sp.* | GM25 | R | AKJQ00000000 |
| *Pseudomonas sp.* | GM49 | R | AKJL00000000 |
| *Pseudomonas sp.* | GM60 | E | AKJI00000000 |
| *Pseudomonas sp.* | GM21 | E | AKJS00000000 |
| *Pseudomonas sp.* | GM55 | E | AKJJ00000000 |
| *Pseudomonas sp.* | GM18 | E | AKJT00000000 |
| *Pseudomonas sp.* | GM48 | R | AKJM00000000 |

[a] E, endosphere; R, rhizosphere.

**Table 1. Summary of genome sequences of 21 rhizosphere and endosphere Pseudomonad species.** Pseudomonad species genome referenced and sequenced as an investigation of the *Populus deltoids* microbiome. Genomes available on NCBI (http://www.ncbi.nlm.nih.gov/genome).

**Genome and PRMT Data**

The genomic data was created using the annotated genome JSON files in KBase. The matrix of genome data was composed of the log base 2 frequency of Enzyme Commission numbers (EC numbers), a numerical classification for enzyme-catalyzed reactions. Thus, bacteria that have enzymes with the same reaction, not necessarily the same genetics, could be grouped together. A PRMT model predicts the enzyme activity and change in turnover of metabolites given the relative abundance of genes for unique enzyme functions **[5]**. The PRMT data was independent of specific nutrient environment and considered secondary metabolism.

**Initial Metabolic Modeling with LB and Carbon-D-Glucose Media**

Annotated genomes of the 21 *Pseudomonas* species were run through the “Build a Metabolic Model” raised on LB Media and Carbon-D-Glucose Media (named Annot_*originalName_media*). Carbon-D-Glucose media contains all of the nutrients necessary for survival, while LB media contains a rich variety of nutrients, not all of which are essential to survival. Each metabolic model was input into a “Run a Flux Balance Analysis” method with the same media used in the metabolic model and renamed (Annot_*originalName_media*_FBA). Next, all 21 flux balance models were inputs in the “Compare FBA Solutions” method. Using the method output of FBA Comparisons, percent similarity of reactions and compounds were each entered into a data table. Homogeneous (R-R, E-E) and heterogeneous (R-E) origin pairings were grouped and averages were computed for each group (R-R, E-E, E-R). In addition, every unique pairing of groups was compared in a Two-Sample T-Test for Equal Variance to determine if there was a statistically significant difference between the reactions occurring in species living in the two media treatments. Since we were trying to differentiate between and eventually predict the origin of the Pseudomonad species, the media with the lower p-value was considered the best candidate for further modification and modeling.

**Media**

Twelve media variations were created using the “Create and Edit Media” method in KBase to represent various differences in the endosphere and rhizosphere to emphasize the difference in the metabolism of the bacteria that inhabit them. Carbon-D-Glucose was kept as a baseline, only containing nutrients essential to survival and D-Glucose, a carbon source used to its maximum exchange flux regardless of bacterial origin. The remaining media varied by their carbon source: the Carbon-D-Glucose media was edited by removing D-Glucose and adding sugars, carbohydrates, organic acids, and amino acids (**Table 2**). The “Comparison FBA Solutions” method output list of reactions showed the D-galacturonate: NAD+ 1-oxidoreductase reaction (rxn01456) was inactive but present exclusively in 12 of our 16 endophytic *Pseudomonas* species. In addition, the Riboflavin transport via proton symport reaction (rxn05645) was noted as active in three of the five rhizosphere inhabiting species and only one endosphere inhabiting species. Thus, two media treatments were created by modifying the Carbon-D-Glucose media by replacing D-Glucose with Riboflavin and D-Galacturonate ($C_6H_{10}O_7$).

| Media Name | Modification to Carbon-D-Glucose Media |
|---|---|
| Carbon-D-Glucose (CDG) | D-Glucose |
| GD | D-Galacturonate |
| LA | L-arabinose |
| MM | Mannose, Proline, Valine |
| Man | Mannose |
| PM | Proline + Mannose |
| VM | Valine + Mannose |
| RM | Riboflavin |
| CM | Oxalate, Citrate, Malonate, Shikimate, Succinate, Propionate, Xylose, Glycine, Ornithine, Alanine, L-cystine, alpha-Cellobiose |
| OGA | Oxalate, Citrate, Malonate, Shikimate, Succinate, Propionate |
| SU | Xylose, alpha-Cellobiose |
| AA | Glycine, Ornithine, Alanine, L-Cystine |

**Table 2. Media Table.** Table of media names and contents. In each media, D-Glucose was removed from the Carbon-D-Glucose Media and other carbon-sources were added. L-arabinose is a compound known to be important to the endophytic lifestyle and oxidized more often by endosphere than rhizosphere *Pseudomonas* species **[7].** A media was constructed by replacing the D-Glucose in the Carbon-D-Glucose Media with L-arabinose to determine if a known endosphere specific media could highlight differences in endosphere and rhizosphere bacteria metabolism. Additionally, it is known that amino acids histidine, proline, valine, alanine, and glycine and carbohydrates glucose, arabinose, mannose, galactose, gluconic acid and xylose are root exudates from rice, key nutrients in the rhizosphere **[8].** Other media produced were constructed of replacing D-Glucose in Carbon-D-Glucose media with Mannose and Valine, Mannose and Proline, and D-Mannose alone. The Complex Media's (CM) contents are metabolites significantly associated with the rhizosphere **[3];** additional media were created composed of CM's added carbon-sources broken down into groups of organic acids, sugars, and amino acids to test if specific groups of compounds helped us differentiate between rhizosphere and endosphere bacteria. The minimum flux exchange had the greatest variance and most often had non-zero values. Since we were trying to emphasize differences in endosphere and rhizosphere bacteria, the minimum uptake exchange flux of each reaction was recorded in our data sets for training our machine learning models.

**FBA Matrices**

Carbon-D-Glucose (minimal media), L-arabinose, D-Galacturonate, Riboflavin, Complex Media (and its subset medias), and the Mixed (Mannose, Proline, Valine) Media were used to build metabolic models and flux balance analyses on all 21 *Pseudomonas* species. For each media treatment, the flux balance analyses were downloaded as excel files and the id of unique reactions and minimum flux exchange each species had with the reaction was stored in an excel sheet to form a matrix with the dimensions of NumberOfUniqueIDReactions x 21 species. For leave-one-out cross validation (LOOCV) purposes in SVM and ANN, any reaction that had the same value across all species or a single species deviation in minimum flux exchange was omitted. In addition to the twelve media-based FBA matrices, an EC matrix was composed of the Enzyme Commission number and the associated log2 of the number of times the annotation appears in the KBase annotated JSON files. A biological context independent metabolic model that takes into account secondary metabolism, Predictive Relative Metabolic Turnover (PRMT), matrix was composed of the relative abundance of enzyme activities across multiple genomes.

## Machine Learning In R

### SVM

Using the e1071 package in R to run support vector machine learning (linear kernel) and LOOCV, we were able to obtain predictions for each of the fourteen matrices. LOOCV, in the context of this experiment, involves keeping 20 of the *Pseudomonas* species as training data and leaving the 21$^{st}$ species' data out to make a prediction and validate. By doing this, we were able to maximize the amount of data the model was trained on and obtain a prediction for every one of the species by changing the training set and validation set.

### ANN

Using the neuralnet package in R to run artificial neural network machine learning and LOOCV, we were able to obtain predictions for each of the fourteen matrices. The ANN was trained on a binomial classification and the following settings: 1 hidden layer, rprop+; 2 hidden layers, backprop; 3 hidden layers, rprop+; 3 hidden layers, backprop. Any settings with the backprop algorithm had a learning rate of 0.01. Since the ANN was trained as supervised machine learning, each species' origin was included in the matrix as 0 or 1, for R and E, respectively. The numerical output, a decimal between 0 and 1, of the validation species' data was rounded to the nearest integer to determine the prediction.

### NMF

Using the NMF package in R to run non-negative matrix factorization machine learning, we were able to obtain basismaps and a set of predictions for each of the fourteen matrices. Since NMF is unsupervised machine learning, the entire matrix with the origin data removed was used in training the model. The original minimum flux exchange numbers did include negative values. These were eliminated by taking the minimum value in the entire data set and shifting all values up by this number, ensuring that the variance was maintained across all of the data. The NMF model was trained with 2, 3, and 4 as the values for basis in order to find the best set of predictions. The numerical categories with the same number for the rhizosphere inhabiting species were assumed to be indicating R and any other numbers were assumed to be E.

## Analysis of Predictions

F1 score (F-score or F-measure) is a form of statistical analysis for binary classification as a measure of accuracy. It is a weighted average of the precision and recall resulting in a number between 1 and 0; 1 indicating higher accuracy and 0 indicating little to no accuracy. The F-score was calculated twice for each set of predictions, once where the rhizosphere origin was considered positive and the second where the endosphere origin was considered positive. F-scores are calculated using the following formula:

Precision is the number of the number of true positive divided by all positive values predicted. Recall is the number of true positive divided by the number of condition positive. If we were considering the rhizosphere positive, we would calculate the number of correctly predicted rhizosphere bacteria, the number predicted rhizosphere bacteria that were actually endosphere bacteria, and the number of predicted endosphere bacteria that were actually rhizosphere bacteria. The f-score of an algorithm that always predicted endosphere was used as a baseline for the endosphere f-score (0.864865). Any model that didn't predict R for any of the species was assigned a 0 for the Rhizosphere f-score. Since we had an unequal number of rhizosphere and endosphere inhabiting *Pseudomonas* species, we prioritized a model that had the highest f-score for rhizosphere and then endosphere.

**Feature Ranking**

Since support vector machine learning does not have an output for sorting important and removing redundant features, there are other packages used to determines such. Using the mlbench and caret packages in R, we were able to first remove redundant features (features that are highly correlated with each other and therefore not important in training a model).
(http://machinelearningmastery.com/feature-selection-with-the-caret-r-package/)

The output of the program is vectors of columns in the data that are highly correlated with each other. Data that are highly correlated and redundant data was removed from the data sets. Secondly, we ranked the features by importance, using a Learning Vector Quantization (LVQ) model. LVQ models are a special kind of artificial neural network that applies a winner-take-all Hebbian learning-based approach. It is a precursor to self-organizing maps (SOM), and related to neural gas, and to the k-Nearest Neighbor algorithm (k-NN). Importance was ranked in a table as well as shown graphically, and the program returned the top 5 most important attributes. Importance was calculated using out-of-bag (OOB) error to measure the error of the data point to the averaged over a forest. OOB estimates help avoid losing information due to a smaller training set, but often underestimate the actual performance improvement and optimal number of iterations. Lastly feature selection was done using Recursive Feature Elimination (RFE). RFE is favored because it makes complex models simpler and easier to interpret, there are shorter training times, and reduces the chance of overfitting by reducing the variance. The graph output after constructing a tree showed the optimal number of variables and performance of the model. After obtaining subsets with the key features, we trained an SVM model on the data subsets of VM and MM to see if we could build a more precise model with the most important features (and determine why these features might be biologically significant).

**Principal Components Analysis**

Principal component analysis (PCA) is a statistical procedure that transforms a set of possibly correlated variables to a set of linearly uncorrelated variables, called principal components. Using the MVN package in R, we loaded the data into a function prcomp() that creates variables that have an equal effect on the ordination (similar to creating a z-score for everything) so that no larger variables dominate the results. To interpret the results, we determined how many principle components to examine (created an abline on the scree plot). In the scree plot we looked for a marked drop in percentage of variation, to determine the important principal components. Then, using a biplot, we graphed various sets of principal components against each other (ie. Pc1 vs. pc2) to look for a visual clustering in rhizosphere and endosphere bacteria.

## Results:

### Genome Annotation

In order to determine whether species origin could be determined by constructing species trees, we created four trees considering 10 (**Figure 1**), 30, 70, 100 neighbors. There were some origin homogeneous sub-clusters, such as the branch including *Pseudomonas* strains GM18, GM79, GM50, GM102, GM67, GM60, GM41, and GM21, suggesting that some sub-clusters may be informative to determining the origin of a species. However, there were also some origin heterogeneous branches where determination of the origin of a species closely related to the heterogeneous branches would be ambiguous. Thus, purely based on the annotated genomic data, a species tree does not contain sufficient information to reliably identify ecological niche.

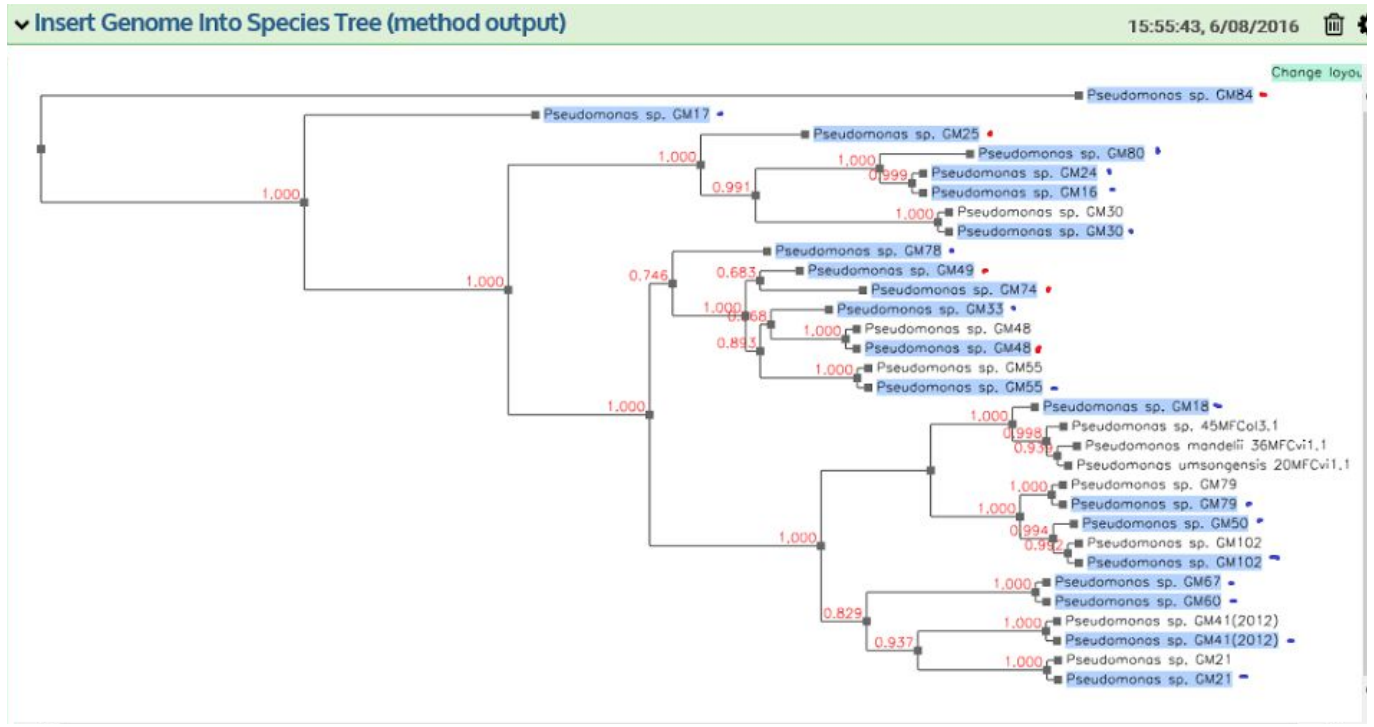

**Figure 1. Species Tree with 10 Neighbors.** Insert Genome Into Species Tree function run on KBase with 10 Neighbors to determine if an identifiable pattern exists genome and origin. Bacteria originating in the rhizosphere are marked in red, endosphere in blue.

**—FOR FULL TABLE, PLEASE CONTACT JENNIFER CHIEN—**
**Table 2. Media-dependent Metabolic Model Matrices**

### Genome Data

There were over 400 unique enzymes (**Table 3**).

**—FOR FULL TABLE, PLEASE CONTACT JENNIFER CHIEN—**
**Table 3. Genome Data (EC numbers)** Log base 2 frequencies of EC numbers for all 21 *Pseudomonas* species.

**PRMT Data**
There were 1778 metabolites (**Table 4**). This data had the greatest number of features and the greatest variability of all data sets used to train the machine learning models.

**—FOR FULL TABLE, PLEASE CONTACT JENNIFER CHIEN—**
**Table 4. PRMT Data.** A positive PRMT score indicates predicted increased metabolic turnover; a negative PRMT score indicates relative greater accumulation of a metabolite. PRMT values are unit-less and do not predict net consumption or production of a metabolite.

**FBA result**

Initial metabolic modeling of all 21 *Pseudomonas* species showed that the reaction and compound similarity in the flux balance models built on Carbon-D-Glucose and LB Media between homogeneous and heterogeneous pairs of bacteria. Although the averages were numerically close, the Carbon-D-Glucose media FBA comparison showed E-E pairings had statistically significantly lower average similarity than R-R pairings (t-test, df=19, p-value < 0.01). The LB media FBA reaction comparison did not have statistically significant differences between any group pairs but the compound similarity did have statistically significant differences. In LB media FBA compound comparison, E-E pairings had statistically significant lower average similarity than R-R pairings (t-test, df=19, p-value < 0.05) (**Table 5**).

| Media | Reaction/ Compound | Pairing | Average | Group | T-Test p-value |
|---|---|---|---|---|---|
| Carbon-D-Glucose | R | R-R | 0.976 | R-R, E-R | 0.064599 |
| | | E-R | 0.97 | E-R, E-E | 0.037595 |
| | | E-E | 0.967 | R-R, E-E | 0.007945 |
| | C | R-R | 1 | R-R, E-R | 1 |
| | | E-R | 1 | E-R, E-E | 1 |
| | | E-E | 1 | R-R, E-E | 1 |
| LB | R | R-R | 0.975 | R-R, E-R | 0.103281 |
| | | E-R | 0.969625 | E-R, E-E | 0.65174 |
| | | E-E | 0.969 | R-R, E-E | 0.06556 |
| | C | R-R | 1 | R-R, E-R | 0.009116 |
| | | E-R | 0.99 | E-R, E-E | 0.503842 |
| | | E-E | 0.98875 | R-R, E-E | 0.010404 |

**Table 5. Flux Balance Analysis Comparison.** Averages were computed for the percent similarity of E-R, R-R, and E-E comparisons. T-tests were conducted and p-values were calculated for all combinations of heterogeneous and homogenous pairings. A p-value of 1 indicates 100% similarity in the averages between the two groups being compared.

The number of unique reactions listed from greatest to least by media name was 143 for Amino Acids, Complex Media, Organic Acids, Proline Media, Sugars, Valine Media; 142 for Mannose Media; 54 for L-arabinose; 53 for D-Galacturonate; 50 for Mixed Media; 23 for Riboflavin; and 16 for Carbon-D-Glucose (after excluding reactions with no more than one species with a different minimum flux exchange value). ▫

**Machine Learning Results**

Twelve matrices constructed from the output of the minimum flux exchange of media based metabolic models (**Table 2**) were used in training SVM, ANN and NMF models. In addition, a PRMT matrix and genome matrix were used to build models with all three machine learning algorithms predict and categorize the 21 *Pseudomonas* bacterial species. For each matrix, the machine learning algorithm settings with the best f-score were recorded (**Table 6)**.

| | | SVM | | ANN | | | | NMF | | |
|---|---|---|---|---|---|---|---|---|---|---|
| | | Rhizo | Endo | Rhizo | Endo | Hidden Layers | Algorithm | Rhizo | Endo | Basis |
| **Genome** | | 0.571 | 0.914 | 0.5 | 0.88 | 1 | rprop+ | 0.545 | 0.839 | 3 |
| **PRMT** | | 0.571 | 0.914 | 0.615 | 0.83 | 2, 3 | backprop | 0.471 | 0.64 | 2 |
| **Carbon-D-Glucose (minimal media)** | | 0 | 0.6 | 0.286 | 0.86 | 1 | rprop+ | 0.667 | 0.867 | 3 |
| **D-Galacturonate** | | 0.545 | 0.839 | 0 | 0.86 | 1 | rprop+ | 0.571 | 0.786 | 2 |
| **L-arabinose** | | 0.6 | 0.875 | 0.333 | 0.89 | 3 | backprop | 0.429 | 0.714 | 2 |
| **Mixed Media [m]** | | 0.8 | 0.938 | 0.286 | 0.86 | 2, 3 | backprop | 0.421 | 0.522 | 3 |
| | **Mannose** | 0.667 | 0.909 | 0.286 | 0.86 | 1 | rprop+ | 0.545 | 0.839 | 3 |
| | **Proline + Mannose** | 0.667 | 0.909 | 0.333 | 0.89 | 3 | backprop | 0.571 | 0.786 | 2 |
| | **Valine + Mannose** | 0.6 | 0.875 | 0.667 | 0.91 | 1, 3 | backprop, rprop+ | 0.429 | 0.714 | 3 |
| **Riboflavin** | | 0.444 | 0.848 | 0.333 | 0.89 | 3 | backprop | 0.571 | 0.786 | 2 |
| **Complex Media [c]** | | 0.286 | 0.643 | 0.333 | 0.89 | 3 | backprop | 0.545 | 0.839 | 2 |
| | **Organic Acids [o]** | 0.6 | 0.875 | 0.333 | 0.89 | 3 | backprop | 0.571 | 0.786 | 2 |
| | **Sugars *** | 0.6 | 0.875 | 0 | 0.86 | 1, 2, 3 | backprop, rprop+ | 0.571 | 0.786 | 2 |
| | **Amino Acids [a]** | 0.4 | 0.813 | 0.333 | 0.89 | 3 | backprop | 0.533 | 0.741 | 2 |

[m] Mannose, Proline, Valine

[c] Oxalate, Citrate, Malonate, Shikimate, Succinate, Propionate, Xylose, Glycine, Ornithine, Alanine, L-cystine, alpha-Cellobiose

[o] Oxalate, Citrate, Malonate, Shikimate, Succinate, Propionate

* Xylose, alpha-Cellobiose

[a] Glycine, Ornithine, Alanine, L-Cystine

**Table 6. Summary Table of SVM, ANN, and NMF f-scores for endosphere and rhizosphere.** Two f-scores were calculated on 14 matrices (genome, PRMT, and 12 media-dependent FBA) where Endosphere and Rhizosphere each were considered the positive. The most predictive model for each machine learning algorithm is in bold. A model that did not correctly predict the origin of any Rhizosphere bacteria was assigned an f-score of 0. The f-score of an algorithm that always predicted endosphere was used as a baseline for the endosphere f-score (0.864865).

The feature selection and PCA results were inconclusive and did not show significant reactions that were essential to predicting endosphere and rhizosphere bacteria. Only one subset of the MM media data had the same precision (f-score) as training the SVM model on the entire MM data set. In addition, all combinations of significant principal components (1-6) were unsuccessful at clearly distinguishing rhizosphere and endosphere bacteria (**Figure 7**).

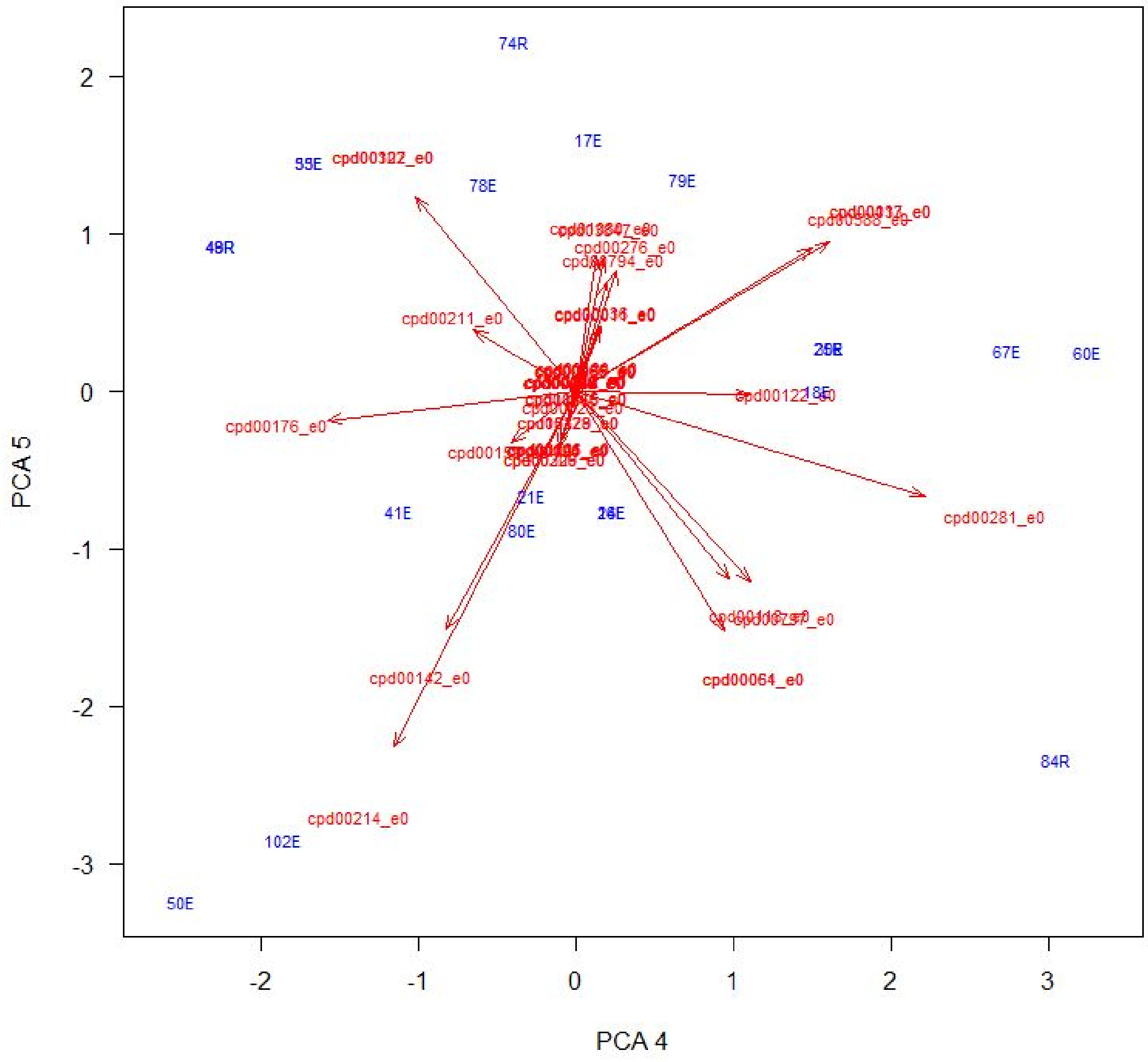

**Figure 7. PCA map of PCA 4 vs PCA 5.** There is no distinguishable cluster of rhizosphere or endosphere bacteria.

## Discussion:

The SVM algorithm had the greatest range in f-scores across all matrices. In addition, the SVM model trained on Mixed Media based FBA matrix had the best f-score of all machine learning models and media, with a rhizosphere f-score of 0.8 and endosphere f-score of  0.9375. The greatest rhizosphere f-scores for ANN and NMF were 0.667, for Valine and Carbon-D-Glucose Media, respectively. The Valine based FBA had an f-score of 0.909, the highest f-score of the ANN models. The Carbon-D-Glucose based FBA had the highest endosphere f-score of all the NMF models with a value of 0.867. The media matrices with the worst predictions for rhizosphere and endosphere bacterial origin were Carbon-D-Glucose for SVM, D-Galacturonate and Sugars for ANN, and Mixed Media for NMF. Since our group of *Pseudomonas* species contained an unequal proportion of rhizosphere and endosphere bacteria, a higher rhizosphere f-score was prioritized.

There was no single media that was the most predictive across all machine-learning algorithms, indicating that there is no single function or gene that defines bacterial niche limitations. In all simulations, the media dependent metabolic models were more predictive than pure genome or media dependent data trained machine learning models. Looking at the metabolic models quantitatively, the endosphere bacteria had more reactions than the rhizosphere bacteria, suggesting the difference between the ability for bacterial inhabitance of either the rhizosphere or endosphere could be the ability to exploit of another niche, rather than bacteria specialized to one specific ecological niche.

The feature ranking approach was not the most effective in this instance; however, it was used to produce a list of important or essential metabolites that can be used in informing future research on *Pseudomonas* bacteria. In addition, although the PCA maps were not effective in grouping the endosphere and rhizosphere bacteria, it can also be used to account for more sophisticated clustering and associations in large data sets.

Metabolic modeling data produced from sequenced and annotated genomes proved to be more informative when using machine learning to understand the interactions between bacteria and their environment. From just twenty-one genomes, we were able to use a more holistic approach to analyze the interaction between plant and bacteria. KBase and machine learning (and possibly feature ranking) can be used to model environments that are not easily reproduced within a lab setting, such as deep-sea micro biomes. In addition, the metabolic modeling and machine learning approach can help isolate and identify multiple key enzymes or reactions essential to differentiating the evolution of key differences between various species. It is possible that a larger set of genomes could allow for a stronger training data set and better prediction values. However, given the current set-up of KBase as a tool, it was most efficient to start with a smaller subgroup to see if the various machine learning approaches used were at all more predictive with metabolic output data than strictly genomic data. Overall, machine learning on metabolic output data proved to be a useful tool for predicting the ecological niches of various *Pseudomonas* species and could be used to simulate and extrapolate information about the complexities of relationships between other organism to their respective environments.

**List of abbreviations:**

ANN – Artificial Neural Network

CDG – Carbon-D-Glucose

FBA – Flux Balance Analysis

NMF – Non-negative Matrix Factorization

PRMT – Predicted Relative Metabolic Turnover

SVM – Support Vector Machine